\documentclass[journal]{IEEEtran}

\usepackage{multirow}
\usepackage{xcolor}
\usepackage[inline]{enumitem}

\usepackage{cite}
\usepackage[bookmarks=false]{hyperref}

\usepackage[pdftex]{graphicx}
\graphicspath{{figures/}}

\usepackage{amsmath}
\usepackage{amssymb}

\DeclareMathOperator*{\argmax}{argmax}
\DeclareMathOperator*{\argmin}{argmin}

\usepackage{array}
\usepackage{stfloats}
\usepackage{url}

\newcommand{\blap}[1]{\smash[b]{\begin{tabular}[t]{@{}c@{}}
                                    #1
\end{tabular}}}

\begin{document}

    \title{Zero-Shot Audio Classification via Semantic Embeddings}

    \author{Huang~Xie,~\IEEEmembership{Student~Member,~IEEE,}
    Tuomas~Virtanen,~\IEEEmembership{Senior~Member,~IEEE}
    \thanks{Manuscript created July 16, 2020; revised November 19, 2020 and February 11, 2020. This work was supported by grants from the European Research Council under the European Union’s H2020 Framework Program through ERC Grant Agreement 637422 EVERYSOUND. (\emph{Corresponding author: Huang Xie}).}
    \thanks{The results of this manuscript have been partly presented in the MSc. thesis \emph{Zero-Shot Learning for Audio Classification} of Huang Xie (accepted in November 2020), but previously not published in a peer-reviewed publication.}
    \thanks{H. Xie and T. Virtanen are with the Faculty of Information Technology and Communication Sciences, Tampere University, Tampere, 33720, Finland (e-mail: huang.xie@tuni.fi; tuomas.virtanen@tuni.fi).}}

    \maketitle

    \IEEEpubid{\begin{minipage}{\textwidth}
                   \ \\[12pt] \centering
                   \copyright 2021 IEEE. Personal use of this material is permitted. Permission from IEEE must be obtained for all other uses, in any current or future media, including reprinting/republishing this material for advertising or promotional purposes, creating new collective works, for resale or redistribution to servers or lists, or reuse of any copyrighted component of this work in other works.
    \end{minipage}}

    \begin{abstract}
        In this paper, we study zero-shot learning in audio classification via semantic embeddings extracted from textual labels and sentence descriptions of sound classes.
        Our goal is to obtain a classifier that is capable of recognizing audio instances of sound classes that have no available training samples, but only semantic side information.
        We employ a bilinear compatibility framework to learn an acoustic-semantic projection between intermediate-level representations of audio instances and sound classes, i.e., acoustic embeddings and semantic embeddings.
        We use VGGish to extract deep acoustic embeddings from audio clips, and pre-trained language models (Word2Vec, GloVe, BERT) to generate either label embeddings from textual labels or sentence embeddings from sentence descriptions of sound classes.
        Audio classification is performed by a linear compatibility function that measures how compatible an acoustic embedding and a semantic embedding are.
        We evaluate the proposed method on a small balanced dataset \mbox{ESC-50} and a large-scale unbalanced audio subset of AudioSet.
        The experimental results show that classification performance is significantly improved by involving sound classes that are semantically close to the test classes in training.
        Meanwhile, we demonstrate that both label embeddings and sentence embeddings are useful for zero-shot learning.
        Classification performance is improved by concatenating label/sentence embeddings generated with different language models.
        With their hybrid concatenations, the results are improved further.
    \end{abstract}


    \IEEEpeerreviewmaketitle

    \section{Introduction}
    \label{sec:introduction}

    \IEEEPARstart{Z}{ero-shot} learning (ZSL), which was first coined in\cite{refs:Palatucci2009Zero}, refers to problems that aim at recognizing instances of classes that have no available training samples but only class side information (e.g., textual descriptions).
    In contrast to supervised classification, samples from only predefined classes are used to train a classifier while new classes get involved at testing or usage stage.
    In this paper, these new classes are referred to as the zero-shot classes.\IEEEpubidadjcol

    Supervised learning has been well-studied for audio classification for decades.
    To obtain classifiers with satisfactory performance, conventional supervised learning methods require large amounts of annotated training samples from target sound classes.
    Due to the high cost of data collection and manual annotation, most of the existing audio datasets\cite{refs:AudioDatasets_online} have limited numbers of audio samples and sound classes, such as \mbox{ESC-50}\cite{refs:Piczak2015ESC} (2,000 labeled audio clips from 50 environmental sound classes) and UrbanSound8K\cite{refs:Salamon2014A} (8,732 labeled audio clips from 10 urban sound classes).
    In recent years, several research works have been conducted to construct large audio datasets with increased number of sound classes, including AudioSet\cite{refs:Gemmeke2017AudioSet} (over 2 million weakly labeled audio clips covering 527 sound classes), Freesound Dataset\cite{refs:Fonseca2017Freesound} (over 290,000 labeled audio clips covering 632 sound classes).
    However, with the increasing number of observed sound classes, it becomes even more challenging for humans to manually collect sufficient annotated samples for all possible sound classes.
    To tackle the lack of adequate training data, recent works\cite{refs:Salamon2017Deep, refs:Koluguri2020Meta, refs:Shi2020FewShot, refs:Chou2019Learning} in the audio literature mainly apply data augmentation\cite{refs:McFee2015ASoftware}, meta learning\cite{refs:Finn2017Model} and few-shot learning\cite{refs:Wang2020Generalizing} methods.
    However, a considerable amount of representative training samples from target classes is still indispensable to make these methods work.
    Furthermore, to classify audio instances from new sound classes, supervised learning classifiers require retraining and exhaustive parameter tuning for the new sound classes, which can be time-consuming.
    With the emergence of zero-shot learning techniques, there is the potential to develop audio classifiers for new sound classes with existing audio datasets\cite{refs:Palatucci2009Zero, refs:Piczak2015ESC, refs:Salamon2014A, refs:Gemmeke2017AudioSet}.\IEEEpubidadjcol

    In recent years, zero-shot learning has received increasing attention in computer vision.
    The key idea of zero-shot learning is to transfer knowledge from training classes to zero-shot classes.
    Due to the lack of training samples from zero-shot classes, side information (e.g., textual descriptions) is required for exploring relationship between training and zero-shot classes to make zero-shot learning possible.
    \mbox{Fu et al.}\cite{refs:Fu2018Recent} reviewed the commonly used side information of visual classes for zero-shot recognition by categorizing them into two sets: semantic attributes and beyond.
    Semantic attributes refer to the intrinsic properties of classes, such as human-defined attributes\cite{refs:Lampert2014Attribute, refs:RomeraParedes2015An}.
    Side information beyond semantic attributes includes concept ontology\cite{refs:Fergus2010Semantic}, and textual descriptions\cite{refs:Akata2016Label, refs:Xian2016Latent}, etc.
    Based on this side information, both training and zero-shot classes can be projected into the same representation space, which enables knowledge transfer between training and zero-shot classes.
    \mbox{Xian et al.}\cite{refs:Xian2018ZeroShot} reviewed three main kinds of methods for zero-shot image recognition.
    A simple method\cite{refs:Lampert2014Attribute} was to learn individual classifiers for independent visual attributes and perform zero-shot classification by combining the predicted attributes of the learned individual classifiers.
    Some sophisticated methods\cite{refs:Akata2016Label, refs:Xian2016Latent} were developed by leveraging intermediate-level representations for images and classes.
    In these methods, a linear/non-linear compatibility function was learned to associate images with classes through their intermediate-level representations.
    For instance, \mbox{Akata et al.}\cite{refs:Akata2016Label} proposed a bilinear compatibility framework to associate images with classes by learning a linear compatibility function.
    \mbox{Xian et al.}\cite{refs:Xian2016Latent} developed this method to explore non-linear compatibilities between images and classes.
    The other methods\cite{refs:Zhang2015ZeroShot, refs:Changpinyo2016Synthesized}, mentioned as hybrid models, fall between independent attribute classifiers and compatibility learning frameworks.

    In contrast to the steady growth of zero-shot learning methods in computer vision, only limited work has been done in the audio field.
    Prior works\cite{refs:Islam2019SoundSemantics, refs:Choi2019ZeroShot, refs:Xie2019ZeroShot} studied compatibility functions for zero-shot learning in audio classification.
    They mapped audio signals into a low-dimensional acoustic space through an acoustic embedding learning module, such as siamese network\cite{refs:Koch2015Siamese} or VGGish\cite{refs:Hershey2017CNN}.
    Sound classes were represented by word embeddings in a semantic space, which were learned from their semantic side information (e.g., textual labels) with pre-trained language embedding models, such as Word2Vec\cite{refs:Mikolov2013Efficient}.
    Then, a compatibility learning module was used to associate acoustic embeddings with semantic embeddings.
    For instance, \mbox{Islam et al.}\cite{refs:Islam2019SoundSemantics} employed a two-layer fully-connected neural network to model a nonlinear compatibility function.
    In our previous work\cite{refs:Xie2019ZeroShot}, we adapted the bilinear compatibility framework\cite{refs:Akata2016Label} from computer vision for zero-shot learning in audio classification.
    Compared with\cite{refs:Islam2019SoundSemantics, refs:Xie2019ZeroShot}, \mbox{Choi et al.}\cite{refs:Choi2019ZeroShot} integrated the acoustic embedding learning module into the compatibility learning module to optimize them holistically.
    Thus, a nonlinear compatibility function was inherently built into their method.

    The previous studies\cite{refs:Islam2019SoundSemantics, refs:Xie2019ZeroShot} used small audio datasets with 2\textendash10 zero-shot classes for evaluation, which would lead to a limited evaluation of zero-shot learning in audio classification.
    In contrast, \mbox{Choi et al.}\cite{refs:Choi2019ZeroShot} studied zero-shot learning for music classification and tagging with large music datasets.
    As a complement, it would be necessary to investigate zero-shot learning in large-scale general audio classification.
    Secondly, only textual labels consisting of 1\textendash3 words were used as the semantic side information of sound classes in\cite{refs:Islam2019SoundSemantics, refs:Choi2019ZeroShot, refs:Xie2019ZeroShot}.
    Compared with textual labels, long pieces of textual descriptions (e.g., sentence descriptions and documents) would convey richer semantic information, which could be useful for improving zero-shot learning performance.
    Thirdly, a pre-trained acoustic embedding model was used for generating acoustic embeddings of audio instances in\cite{refs:Xie2019ZeroShot}.
    Since a pre-trained acoustic embedding model could have already embedded acoustic information about zero-shot classes, using it to generate acoustic embeddings at training stage would lead to a biased evaluation of zero-shot learning.

    In this paper, the present study is extended from our previous work\cite{refs:Xie2019ZeroShot}.
    The main contributions of this work are the following:
    \begin{itemize}
        \item We evaluate zero-shot learning in-depth on both a small environmental dataset \mbox{ESC-50}\cite{refs:Piczak2015ESC} and a large general dataset AudioSet\cite{refs:Gemmeke2017AudioSet}.
        \item We consider textual labels and additional sentence descriptions as the side information of sound classes for zero-shot learning in audio classification.
        Three kinds of semantic representations of sound classes are studied: label embeddings, sentence embeddings, and their concatenations.
        \item We investigate generating semantic embeddings with different pre-trained language models: Word2Vec\cite{refs:Mikolov2013Efficient}, GloVe\cite{refs:Pennington2014GloVe}, and BERT\cite{refs:Devlin2019BERT}.
    \end{itemize}

    The remainder of this paper is organized as follows.
    Section~\ref{sec:proposed-method} presents an overview of the proposed method for zero-shot learning in audio classification.
    Section~\ref{sec:intermediate-level-representations} introduces the implementation details of generating acoustic embeddings and semantic embeddings.
    Section~\ref{sec:bilinear-compatibility-learning-framework} describes the bilinear compatibility learning framework used for associating acoustic embeddings with semantic embeddings.
    Section~\ref{sec:experiments} describes the evaluation experiments, and Section~\ref{sec:results-and-analysis} discusses the results.
    Section~\ref{sec:conclusion} concludes this paper.

    \section{Proposed Method}
    \label{sec:proposed-method}

\begin{figure*}[!t]
        \centering
        \includegraphics[width=0.7\textwidth]{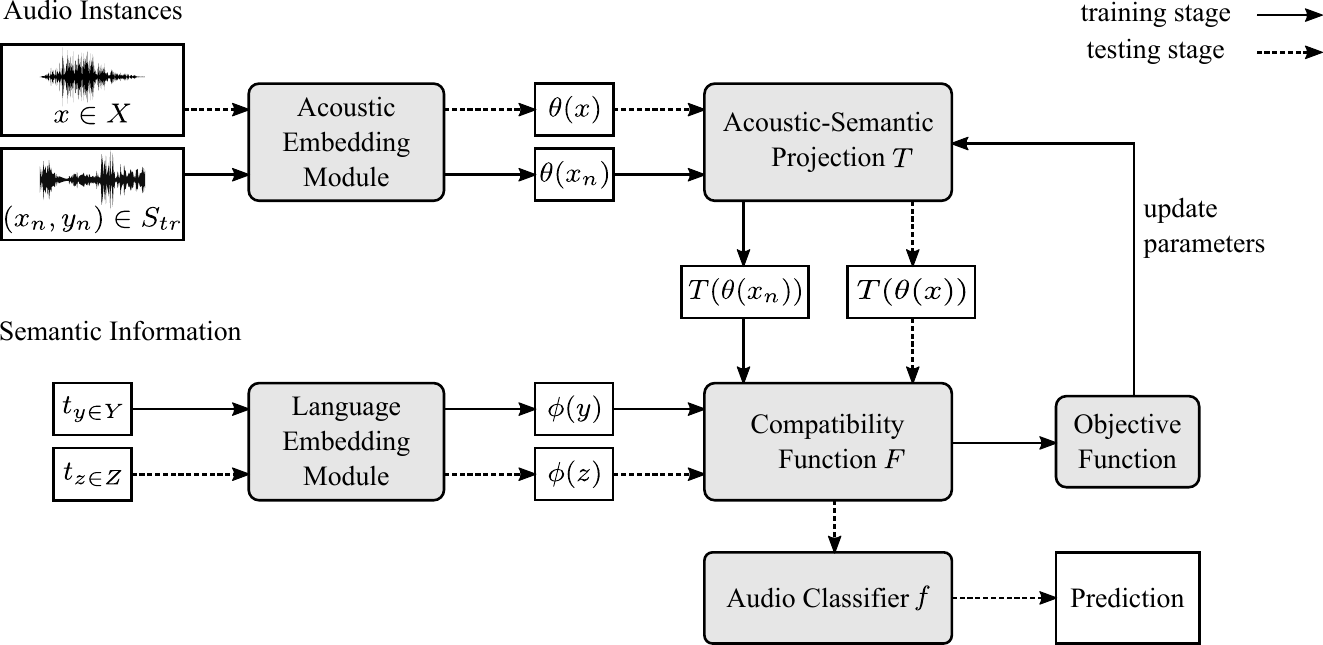}
        \caption{Zero-shot learning via semantic information in audio classification: sound classes $y \in Y$ and $z \in Z$ are represented by semantic embeddings extracted from their respective semantic side information $t_{y}$ and $t_{z}$. 1) At training stage (solid line), an acoustic-semantic projection $T$ is optimized with every audio sample $(x_{n},y_{n}) \in S_{tr}$. 2) At testing stage (dashed line), an audio classifier $f$ will classify an audio instance $x \in X$ into a sound class $z \in Z$ that has the maximal compatibility value $F(T(\theta(x)),\phi(z))$.}
        \label{fig:zsl_training_testing}
    \end{figure*}

    In this section, we present an overview of the proposed method for zero-shot learning in audio classification, as illustrated in~\figurename{\ref{fig:zsl_training_testing}}.
    Sound classes are represented by semantic embeddings, which are learned from their semantic side information with a language embedding module.
    Acoustic embeddings are extracted from audio instances through an acoustic embedding module, and then projected onto semantic embeddings through an acoustic-semantic projection.
    A compatibility function is defined to measure how similar/compatible an projected acoustic embedding and a semantic embedding are.
    During training, the acoustic-semantic projection is optimized with training data.
    For prediction, an audio instance is classified into a sound class, the semantic embedding of which has the maximal compatibility with its projected acoustic embedding.

    We denote by $X$ an audio sample space, $Y$ a set of training classes, and $Z$ a set of zero-shot classes.
    Note that $Y \cap Z = \emptyset$.
    We define $t_{y}$, $t_{z}$ as the semantic side information of sound classes $y \in Y$ and $z \in Z$, respectively.
    We are given a set of training samples $S_{tr}=\{(x_{n},y_{n}) \in X \times Y | n=1,\dots,N\}$, where $x_{n}$ is an annotated audio sample belonging to a training class $y_{n}$.
    Our goal is to obtain an audio classifier $f:X \rightarrow Z$ that can recognize zero-shot classes $Z$, which have no available training samples but only semantic side information.
    In the proposed method, zero-shot learning is done by leveraging intermediate-level representations of audio instances and sound classes, i.e., acoustic embeddings and semantic embeddings.
    We denote by $\theta(x) \in \mathbb{R}^{d_{a}}$ the acoustic embedding of an audio instance $x \in X$.
    We denote by $\phi(y) \in \mathbb{R}^{d_{s}}$, $\phi(z) \in \mathbb{R}^{d_{s}}$ the semantic embedding extracted from semantic side information $t_{y}$ and $t_{z}$, respectively.
    With these embeddings, an acoustic-semantic projection $T:\mathbb{R}^{d_{a}} \rightarrow \mathbb{R}^{d_{s}}$ is exploited to associate audio instances with sound classes.

    Given an audio instance $x \in X$ belonging to a sound class $z \in Z$, we assume that the projected acoustic embedding $T(\theta(x))$ in the semantic embedding space is more similar to the semantic embedding $\phi(z)$ rather than those of other sound classes.
    A similarity scoring function $F:\mathbb{R}^{d_{s}} \times \mathbb{R}^{d_{s}} \rightarrow \mathbb{R}$, as known as the compatibility function, is then defined to measure how similar/compatible an projected acoustic embedding and a semantic embedding are.
    Possible choices of $F$ could be Euclidean distance\cite{refs:Islam2019SoundSemantics}, cosine similarity\cite{refs:Choi2019ZeroShot}, etc.
    Therefore, the audio classifier $f:X \rightarrow Z$ is formulated as\footnote{Note that the $\argmin$ operation should be considered for other compatibility functions, such as Euclidean distance, etc.}
    \begin{equation}
        \label{eq:ZSL_classifier}
        f(x) = \argmax_{z \in Z} F(T(\theta(x)),\phi(z)).
    \end{equation}
    At training stage, $f$ is trained with every audio sample $(x_{n},y_{n}) \in S_{tr}$, and is expected to generalize to new classes $Z$ as in~\eqref{eq:ZSL_classifier} at testing stage.

    \section{Intermediate-Level Representations}
    \label{sec:intermediate-level-representations}

    In this section, we describe our approaches of generating intermediate-level representations of audio instances and sound classes, i.e., acoustic embeddings and semantic embeddings.

    \subsection{Semantic Embeddings}
    \label{subsec:semantic-embeddings}

    In this work, we use three pre-trained language embedding models to extract semantic embeddings from semantic side information of sound classes.
    Specifically, we adopt Word2Vec\cite{refs:Mikolov2013Efficient} and GloVe\cite{refs:Pennington2014GloVe} as word embedding models to tackle textual labels, and we employ BERT\cite{refs:Devlin2019BERT} as a sentence embedding model to process sentence descriptions.

    Word2Vec (Skip-gram model) is a two-layer fully-connected neural network, which is capable of learning word embeddings that are useful for predicting surrounding words in a sentence or a document\cite{refs:Mikolov2013Efficient, refs:Mikolov2013Distributed}.
    For the sake of simplicity, we use a publicly available pre-trained Word2Vec\cite{refs:Word2Vec_online}, which is trained on Google News dataset.
    It consists of 300-dimensional word embeddings for 3 million case-sensitive English words and phrases.
    To represent sound classes, we take the average of word embeddings of individual words/phrases contained in their textual labels.

    GloVe is a statistical language model, which learns word embeddings based on the statistical information of global word-word co-occurrence and local context of words\cite{refs:Pennington2014GloVe}.
    We adopt a publicly available pre-trained GloVe\cite{refs:GloVe_online}, which is trained on documents from Common Crawl dataset.
    It contains 300-dimensional word embeddings for roughly 2.2 million case-sensitive English words.
    Similarly, we average GloVe word embeddings to obtain semantic embeddings for sound classes.

    BERT\cite{refs:Devlin2019BERT}, which stands for Bidirectional Encoder Representations from Transformers, is a contextual language understanding model that is capable of learning deep semantic embeddings of contiguous texts, such as sentences and documents.
    It obtains state-of-the-art results on a broad set of natural language processing tasks\cite{refs:Devlin2019BERT}.
    In this work, we use a pre-trained BERT model\cite{refs:BERT_online} to produce semantic embeddings of sound classes from their sentence descriptions.
    The pre-trained model consists of 24 layers, 1024 hidden states and 16 heads, and it is trained on a large lowercased text corpus (Wikipedia + BookCorpus).
    At usage stage, it outputs a 1024-dimensional embedding for a sentence description.

    \subsection{Acoustic Embeddings}
    \label{subsec:acoustic-embeddings}

    In this work, we use VGGish\cite{refs:Hershey2017CNN} to generate acoustic embeddings from audio clips.
    VGGish\cite{refs:Hershey2017CNN} is a convolutional neural network, which is derived from an VGGNet with \mbox{Configuration A}\cite{refs:Simonyan2015Very} by adopting the following changes:
    \begin{itemize}
        \item The input size is changed to $96\times64$ for audio log mel spectrograms.
        \item The last group of convolutional and max-pool layers is dropped.
        \item The last fully connected layer is replaced with a 128-wide fully connected layer.
    \end{itemize}
    The authors in\cite{refs:Hershey2017CNN} used VGGish to perform feature extraction for large-scale audio classification tasks and defined the output embeddings of the last 128-wide fully connected layer as acoustic features.
    It was shown that audio classification performance was improved by using these acoustic embeddings instead of traditional hand-crafted features (e.g., log mel spectrogram)\cite{refs:Hershey2017CNN}.

    In zero-shot learning, it is assumed that there is no available acoustic data from zero-shot classes at training stage.
    In\cite{refs:Xie2019ZeroShot}, we extracted acoustic embeddings from audio clips with a pre-trained VGGish.
    We realize that the pre-trained VGGish may have already embedded acoustic information about zero-shot classes and using it to generate acoustic embeddings at training stage can lead to a biased evaluation of zero-shot learning.
    In this work, we train VGGish from scratch and ensure that test classes are excluded from training data.

    To obtain acoustic embeddings with VGGish, we follow the processing in\cite{refs:Hershey2017CNN}.
    At first, all the audio clips are resampled to 16 kHz mono.
    Each mono clip is split into non-overlapping segments of 960 ms.
    For each segment, short-time Fourier transform is then computed on 25 ms Hanning windowed frames with a step size of 10 ms.
    After that, the power spectrogram is aggregated into 64 mel bands.
    Finally, a $96\times64$ log mel spectrogram is generated and fed into VGGish to obtain a 128-dimensional embedding vector for every audio segment.
    To generate a clip-level acoustic embedding for an audio clip, we simply average the 128-dimensional embedding vectors of its 960 ms segments.

    \section{Bilinear Compatibility Learning Framework}
    \label{sec:bilinear-compatibility-learning-framework}

    In this section, we describe the bilinear compatibility learning framework adapted from computer vision\cite{refs:Akata2016Label} to tackle zero-shot learning in audio classification in this work.

    \subsection{Bilinear Compatibility}
    \label{subsec:bilinear-compatibility}

    As illustrated in~\figurename{\ref{fig:zsl_training_testing}}, acoustic embeddings are extracted from audio instances through an acoustic embedding module.
    Similarly, sound classes are represented by semantic embeddings learned from their semantic side information with a language embedding module.
    After that, acoustic embeddings are projected onto semantic embeddings through a simple linear projection $T$:
    \begin{equation}
        \label{eq:acoustic_semantic_proj}
        T(\theta(x)) = W^{T} \theta(x),
    \end{equation}
    where $W$ is the projection matrix to be learned.
    For compatibility function $F$, a natural parameterization is the dot product between $T(\theta(x))$ and $\phi(z)$.
    Therefore, $F$ is writen as
    \begin{equation}
        \label{eq:dot_production}
        F(T(\theta(x)),\phi(z)) = T(\theta(x))^{T} \phi(z).
    \end{equation}
    By substituting~\eqref{eq:acoustic_semantic_proj} into~\eqref{eq:dot_production}, $F$ is reformulated as
    \begin{equation}
        \label{eq:comp_function}
        F(T(\theta(x)),\phi(z)) = \theta(x)^{T} W \phi(z),
    \end{equation}
    which is bilinear with respect to acoustic/semantic embeddings.
    For zero-shot learning, the audio classifier $f:X \rightarrow Z$ is then defined as
    \begin{equation}
        \label{eq:bilinear_model}
        f(x) = \argmax_{z \in Z} \theta(x)^{T} W \phi(z).
    \end{equation}

    \subsection{Training Algorithm}
    \label{subsec:training-algorithm}

    In this section, we introduce the algorithm for optimizing $W$ in~\eqref{eq:acoustic_semantic_proj} with training data $S_{tr}$.
    Given an audio sample $(x_{n},y_{n}) \in S_{tr}$, we consider the task of sorting sound classes $y \in Y$ in descending order according to their compatibility values $F(T(\theta(x_{n})),\phi(y))$.
    Our objective is to optimize $W$ so that $y_{n}$ will be ranked at top of the sorted class list, i.e., having the maximal compatibility value for $x_{n}$.

    Let $r_{y_{n}}$ be the rank of sound class $y_{n}$, which refers to the number of incorrect classes placed before $y_{n}$.
    \mbox{Usunier et al.}\cite{refs:Usunier2009Ranking} proposed a ranking error function that transformed rank $r_{y_{n}}$ into loss $\beta(r_{y_{n}})$:
    \begin{equation}
        \label{eq:beta_loss}
        \beta(r_{y_{n}}) = \sum_{i=1}^{r_{y_{n}}} \alpha_{i},
    \end{equation}
    with $\alpha_{1} \geq \alpha_{2} \geq \dots \geq 0$ and $\beta(0)=0$.
    Specifically, $\beta(r_{y_{n}})$ defines a ranking penalty for $y_{n}$, and $\alpha_{i}$ measures the penalty incurred by losing a rank from $i-1$ to $i$.
    In this work, we follow previous work in\cite{refs:Akata2016Label, refs:Usunier2009Ranking} and choose $\alpha_{i}=1/i$.

    Inspired by\cite{refs:Usunier2009Ranking}, \mbox{Weston et al.}\cite{refs:Weston2011WSABIE} introduced the hinge loss $l$ into~\eqref{eq:beta_loss} to add a margin and make it continuous, and then proposed the so-called weighted approximate-rank pairwise loss function
    \begin{equation}
        \label{eq:weighted_ranking_loss}
        \dfrac{1}{N} \sum_{n=1}^{N} \dfrac{\beta(r_{y_{n}})}{r_{y_{n}}} \sum_{y\in Y} \max \{0, l(x_{n},y_{n},y)\},
    \end{equation}
    with the convention $0/0=0$ when $r_{y_{n}}=0$, i.e., $y_{n}$ was top-ranked.
    We define the hinge loss $l(x_{n},y_{n},y)$ as
    \begin{equation}
        \label{eq:hinge_loss}
        \begin{split}
            l(x_{n},y_{n},y) = \Delta(y_{n},y) &+ F(T(\theta(x_{n})),\phi(y)) \\ &- F(T(\theta(x_{n})),\phi(y_{n})),
        \end{split}
    \end{equation}
    where $\Delta(y_{n},y)=0$ if $y_{n}=y$ and 1 otherwise.

    The loss function~\eqref{eq:weighted_ranking_loss} is convex and can be optimized through stochastic gradient descent.
    By minimizing~\eqref{eq:weighted_ranking_loss}, the correct class of an audio sample will be top-ranked, i.e., having the maximal compatibility.
    To prevent over-fitting, we regularize~\eqref{eq:weighted_ranking_loss} with the squared Frobenius norm of $W$.
    The final objective function is
    \begin{equation}
        \label{eq:objective}
        \dfrac{1}{N} \sum_{n=1}^{N} \dfrac{\beta(r_{y_{n}})}{r_{y_{n}}} \sum_{y\in Y} \max \{0, l(x_{n},y_{n},y)\} + \lambda \| W \|^{2},
    \end{equation}
    where $\lambda$ is the coefficient of the regularization term.
    In this work, we select $\lambda$ from $\{0, 0.01, 1, 10\}$ with a validation dataset at training stage.

    \section{Experiments}
    \label{sec:experiments}

    In this section, we describe the evaluation experiments on two audio datasets: \mbox{ESC-50}\cite{refs:Piczak2015ESC} and AudioSet\cite{refs:Gemmeke2017AudioSet}.

    \subsection{Datasets}
    \label{subsec:datasets}

    \mbox{ESC-50}\cite{refs:Piczak2015ESC} is a small balanced audio dataset, which includes 2,000 single-label 5-second audio clips covering 50 environmental sound classes, as shown in Table~\ref{tab:ESC50_class}.
    These sound classes are arranged into five high-level sound categories with 10 classes per category: animal sounds, natural soundscapes \& water sounds, human (non-speech) sounds, interior/domestic sounds, and exterior/urban noises.
    Each class is described using a textual class label, such as ``dog", ``door wood knock".

    AudioSet\cite{refs:Gemmeke2017AudioSet} is an unbalanced large general audio dataset, which contains roughly 2 million multi-label audio clips covering over 527 sound classes.
    For the sake of simplicity, we consider only single-label classification in this work and select single-label audio clips from AudioSet.
    To make a trade-off between least populated classes (i.e., have no more than 50 audio samples) and most populated classes (i.e., have more than 1,000 audio samples), we randomly under-sample the most populated classes until each of them has at most 1,500 audio samples.
    The final audio subset extracted from AudioSet contains 112,774 single-label 10-second audio clips and 521 sound classes.
    Each of these classes is defined by a textual label.
    Meanwhile, AudioSet\cite{refs:Gemmeke2017AudioSet} provides an additional sentence description for every sound class as an explanation of its meaning and characteristics.

    \begin{table}[!t]
        \renewcommand{\arraystretch}{1.3}
        \caption{Sound Category Groups in ESC-50}
        \label{tab:ESC50_class}
        \centering
        \begin{tabular}{c||p{0.6\linewidth}}
            \hline
            \bfseries Sound Category & \bfseries Sound Classes                                                                                                 \\
            \hline\hline
            Animal sounds            & dog, rooster, pig, cow, frog, cat, hen, insects, sheep, crow                                                            \\
            \hline
            Natural sounds           & rain, sea waves, crackling fire, crickets, chirping birds, water drops, wind, pouring water, toilet flush, thunderstorm \\
            \hline
            Human sounds             & crying baby, sneezing, clapping, breathing, coughing, footsteps, laughing, brushing teeth, snoring, drinking sipping    \\
            \hline
            \blap{Interior/domestic \\sounds} & door wood knock, mouse click, keyboard typing, door wood creaks, can opening, washing machine, vacuum cleaner, clock alarm, clock tick, glass breaking \\
            \hline
            \blap{Exterior/urban \\noises} & helicopter, chainsaw, siren, car horn, engine, train, church bells, airplane, fireworks, hand saw \\
            \hline
        \end{tabular}
    \end{table}

    \subsection{Experiments on ESC-50}
    \label{subsec:experiments-on-esc-50}

    In this section, we introduce our experimental setups on \mbox{ESC-50}\cite{refs:Piczak2015ESC}.

    \subsubsection{Dataset Splits}
    We conduct zero-shot learning with 5-fold cross-validation on \mbox{ESC-50} sound classes.
    First, we split sound classes into five disjoint class folds with two partition strategies: based on sound categories and randomly.
    In the category-based strategy, sound classes belonging to the same high-level category are organized as one class fold, which is shown as ``Sound Category'' in Table~\ref{tab:ESC50_class}.
    In the random-based strategy, we group sound classes at random into five class folds, as shown in Table~\ref{tab:ESC50_random_class}.
    For each partition strategy, we apply 5-fold cross-validation.
    Four class folds (40 sound classes with 1,600 audio samples) are used to train the bilinear compatibility framework (in Section~\ref{sec:bilinear-compatibility-learning-framework}).
    After that, zero-shot learning is conducted on the remaining class fold (10 sound classes with 400 audio samples).

    \begin{table}[!t]
        \renewcommand{\arraystretch}{1.3}
        \caption{Random-based Class Folds in ESC-50}
        \label{tab:ESC50_random_class}
        \centering
        \begin{tabular}{c||p{0.6\linewidth}}
            \hline
            \bfseries Class Fold & \bfseries Sound Classes                                                                                                   \\
            \hline\hline
            Fold0                & brushing teeth, church bells, clock tick, cow, drinking sipping, fireworks, helicopter, mouse click, pig, washing machine \\
            \hline
            Fold1                & clapping, crickets, glass breaking, hand saw, keyboard typing, laughing, siren, sneezing, thunderstorm, vacuum cleaner    \\
            \hline
            Fold2                & breathing, chainsaw, chirping birds, coughing, door wood creaks, door wood knock, frog, pouring water, rain, train        \\
            \hline
            Fold3                & airplane, can opening, crying baby, engine, footsteps, hen, insects, rooster, snoring, toilet flush                       \\
            \hline
            Fold4                & car horn, cat, clock alarm, crackling fire, crow, dog, sea waves, sheep, water drops, wind                                \\
            \hline
        \end{tabular}
    \end{table}

    \subsubsection{VGGish Training}
    We train VGGish from scratch with the same class folds that are used for training the bilinear compatibility framework.
    In total, there are 1,600 audio samples of 40 sound classes from four class folds.
    Each audio sample is split into five 960 ms segments annotated with its sound class.
    We obtain 8,000 audio segments, and then randomly divide them into two sets with a class-specific proportion of 80/20: 6,400 segments for training and 1,600 segments for parameter validation.
    During the 5-fold cross-validation, a separate VGGish is trained from scratch at each step, using Adam\cite{refs:Kingma2015Adam} optimizer with a learning rate of 1e$-$4.
    Training is terminated by early stopping, i.e., once performance stops increasing on the validation partition.

    \subsubsection{Class Semantic Embeddings}
    There are only textual labels of sound classes available in \mbox{ESC-50}.
    Hence, we consider zero-shot learning via class label embeddings on \mbox{ESC-50}.
    With the pre-trained Word2Vec\cite{refs:Word2Vec_online} and GloVe\cite{refs:GloVe_online} (in Section~\ref{subsec:semantic-embeddings}), we obtain three sets of class label embeddings: Word2Vec label embeddings (WLE), GloVe label embeddings (GLE), and their concatenations (WLE+GLE).

    \subsection{Experiments on AudioSet}
    \label{subsec:experiments-on-audioset}

    In this section, we introduce our experimental setups on the extracted subset of AudioSet\cite{refs:Gemmeke2017AudioSet}.

    \subsubsection{Dataset Splits}
\begin{figure*}[!t]
        \centering
        \includegraphics[width=\textwidth]{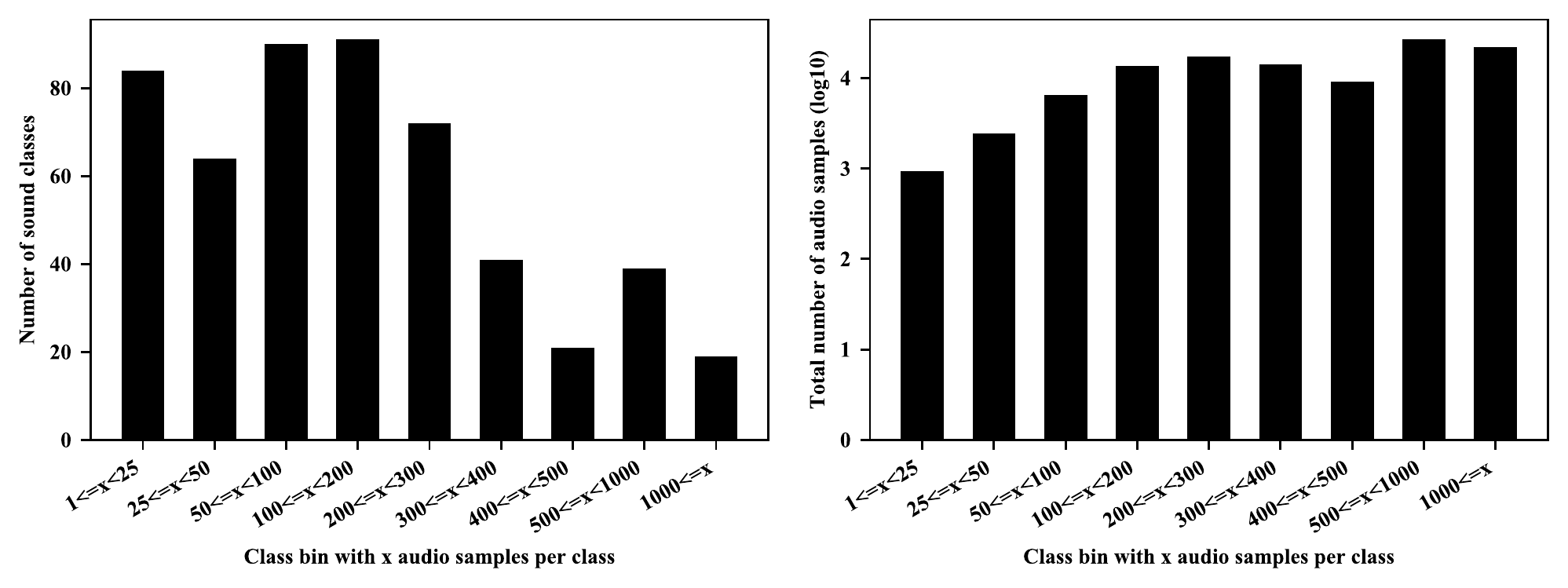}
        \caption{Number of sound classes and total audio samples of class bins in the extracted subset of AudioSet.}
        \label{fig:fig_class_bins}
    \end{figure*}

    We split the extracted subset (in Section~\ref{subsec:datasets}) into five disjoint class folds, which have similar number of sound classes and audio samples, as shown in Table~\ref{tab:AudioSet_class}.
    First, all sound classes are arranged into nine class bins according to their number of audio samples.~\figurename{\ref{fig:fig_class_bins}} illustrates the number of sound classes and total audio samples in the nine class bins.
    Then, each class bin is randomly split into five groups of sound classes with equal size.
    After that, we randomly select one group from each class bin without replacement, and merge the nine selected groups into one class fold.

    With the increased number of sound classes and audio samples in the extracted subset, we explore the effect of training VGGish with different class folds on generating acoustic embeddings for zero-shot learning.
    Table~\ref{tab:AudioSet_data_setting} presents two data settings for training VGGish from scratch.
    In \mbox{Setting 1}, we train an VGGish model with class folds ``Fold0" and ``Fold1", which are excluded from zero-shot learning.
    As a comparison, another VGGish model is trained in \mbox{Setting 2} with class folds ``Fold2" and ``Fold3", which are used for training and validating the bilinear compatibility framework.
    For test, zero-shot learning is conducted on class fold ``Fold4''.

    \begin{table}[!t]
        \renewcommand{\arraystretch}{1.3}
        \caption{Class Folds in the Extracted Subset of AudioSet}
        \label{tab:AudioSet_class}
        \centering
        \begin{tabular}{c||c||c}
            \hline
            \bfseries Class Fold & \bfseries Total Sound Classes & \bfseries Total Audio Samples \\
            \hline\hline
            Fold0                & 104                           & 23007                         \\
            \hline
            Fold1                & 104                           & 22889                         \\
            \hline
            Fold2                & 104                           & 22762                         \\
            \hline
            Fold3                & 104                           & 22739                         \\
            \hline
            Fold4                & 105                           & 21377                         \\
            \hline
        \end{tabular}
    \end{table}
    \begin{table}[!t]
        \renewcommand{\arraystretch}{1.3}
        \caption{Two Data Settings with the Extracted Subset of AudioSet}
        \label{tab:AudioSet_data_setting}
        \centering
        \begin{tabular}{c||c||c}
            \hline
            \bfseries Model              & \bfseries Setting 1 & \bfseries Setting 2 \\
            \hline\hline
            VGGish training + validation & Fold0, Fold1        & Fold2, Fold3        \\
            \hline
            ZSL training & \multicolumn{2}{c}{Fold2} \\
            \hline
            ZSL validation & \multicolumn{2}{c}{Fold3} \\
            \hline
            ZSL test & \multicolumn{2}{c}{Fold4} \\
            \hline
        \end{tabular}
    \end{table}

    \subsubsection{VGGish Training}
    We train an VGGish from scratch with two class folds in each data setting.
    Audio samples are randomly divided into training/validation partitions with a class-specific proportion of 75/25.
    As described in Section~\ref{subsec:acoustic-embeddings}, we split audio clips into non-overlapping 960 ms segments and extract a $96\times64$ log mel spectrogram for each segment.
    Then, log mel spectrograms belonging to the same audio sample is fed into VGGish as a single input batch.
    During training, we use Adam optimizer with a learning rate of 1e$-$4.
    We terminate training by early stopping, i.e., once performance stops increasing on the validation partition.

    \begin{table}[!t]
        \renewcommand{\arraystretch}{1.2}
        \caption{Semantic Embeddings for AudioSet Sound Classes}
        \label{tab:AudioSet_semantic_embedding}
        \centering
        \begin{tabular}{c||c||c}
            \hline
            \bfseries Semantic Embedding & \bfseries Type & \bfseries Dimensionality \\
            \hline\hline
            WLE                          & Label          & 300                      \\
            \hline
            GLE                          & Label          & 300                      \\
            \hline
            WLE+GLE                      & Label          & 600                      \\
            \hline
            WSE                          & Sentence       & 300                      \\
            \hline
            GSE                          & Sentence       & 300                      \\
            \hline
            BSE                          & Sentence       & 1024                     \\
            \hline
            WSE+GSE                      & Sentence       & 600                      \\
            \hline
            WSE+BSE                      & Sentence       & 1324                     \\
            \hline
            GSE+BSE                      & Sentence       & 1324                     \\
            \hline
            WSE+GSE+BSE                  & Sentence       & 1624                     \\
            \hline
            WLE+WSE                      & Hybrid         & 600                      \\
            \hline
            GLE+GSE                      & Hybrid         & 600                      \\
            \hline
            WLE+BSE                      & Hybrid         & 1324                     \\
            \hline
            GLE+BSE                      & Hybrid         & 1324                     \\
            \hline
            WLE+GLE+BSE                  & Hybrid         & 1624                     \\
            \hline
        \end{tabular}
    \end{table}

    \subsubsection{Class Semantic Embeddings}
    In addition to textual labels, AudioSet provides an sentence description for every sound class as an explanation of its meaning and characteristics.
    We consider these sentence descriptions as a complement of the semantic side information of sound classes for zero-shot learning.
    Therefore, sound classes have two types of semantic embeddings: label embeddings, which are extracted from their textual labels, and sentence embeddings, which are generated from their sentence descriptions.
    We obtain three sets of label embeddings (i.e., WLE, GLE and WLE+GLE) by using the pre-trained Word2Vec\cite{refs:Word2Vec_online} and GloVe\cite{refs:GloVe_online} (in Section~\ref{subsec:semantic-embeddings}).
    For sentence embeddings, we first use the pre-trained BERT\cite{refs:Devlin2019BERT} to tackle sentence descriptions and obtain a set of BERT sentence embeddings (BSE).
    As a comparison, we experiment on generating sentence embeddings with the pre-trained Word2Vec\cite{refs:Word2Vec_online} and GloVe\cite{refs:GloVe_online}.
    An Word2Vec/GloVe sentence embedding (WSE/GSE) is defined as the average of Word2Vec/GloVe word embeddings of individual words in a sentence description.
    Since stop words usually contain useless semantic information, we exclude them from calculating WSE/GSE\@.
    Furthermore, we experiment with concatenating semantic embeddings to obtain concatenated sentence embeddings (e.g., WSE+GSE) and hybrid embeddings (e.g., WLE+BSE) as shown in Table~\ref{tab:AudioSet_semantic_embedding}.

    \subsection{Evaluation Metrics}
    \label{subsec:evaluation-metrics}

    For evaluation, we calculate the classification accuracy (Top-1) and mean average precision (mAP) over samples.
    Top-1 is the proportion of correctly classified samples to the total samples.
    mAP is the mean over samples of the average precision (AP), which is the average of precisions at all positions where correct classes are placed in a sorted class list\cite{refs:Buckley2004Retrieval}.
    In this work, mAP is inversely proportional to the ranks of corrected classes (in Section~\ref{subsec:training-algorithm}), and perfect classification achieves a mAP of 1.0 when all correct classes are top-ranked.

    \section{Results and Analysis}
    \label{sec:results-and-analysis}

    In this section, we discuss the experimental results of zero-shot learning via semantic embeddings on \mbox{ESC-50}\cite{refs:Piczak2015ESC} and AudioSet\cite{refs:Gemmeke2017AudioSet}.

    \subsection{Zero-Shot Learning on ESC-50}
    \label{subsec:zero-shot-learning-on-esc-50}

    We report the results of the category-based strategy in Table~\ref{tab:ESC50_category}, and the ones of the random-based strategy in Table~\ref{tab:ESC50_random}, respectively.
    For 5-fold cross-validation, the averaged results of these strategies are compared in Table~\ref{tab:ESC50_average}.

    \begin{table*}[!t]
        \renewcommand{\arraystretch}{1.3}
        \caption{Zero-Shot Classification on ESC-50 with Category Partition Strategy}
        \label{tab:ESC50_category}
        \centering
        \begin{tabular}{c||c|c||c|c||c|c}
            \hline
            \multirow{2}{*}{\bfseries Test Class Fold} & \multicolumn{2}{c||}{\bfseries WLE} & \multicolumn{2}{c||}{\bfseries GLE}
            & \multicolumn{2}{c}{\bfseries WLE+GLE}
            \\
            \cline{2-7}
            & mAP            & Top-1          & mAP  & Top-1 & mAP  & Top-1 \\
            \hline\hline
            Animal sounds            & 0.43           & 0.26           & 0.37 & 0.18  & 0.38 & 0.20  \\
            \hline
            Natural sounds           & \bfseries 0.52 & \bfseries 0.33 & 0.42 & 0.24  & 0.42 & 0.24  \\
            \hline
            Human sounds             & 0.33           & 0.18           & 0.35 & 0.20  & 0.34 & 0.18  \\
            \hline
            Interior/domestic sounds & 0.37           & 0.21           & 0.38 & 0.21  & 0.38 & 0.21  \\
            \hline
            Exterior/urban noises    & 0.38           & 0.19           & 0.40 & 0.21  & 0.39 & 0.21  \\
            \hline
        \end{tabular}
    \end{table*}
    \begin{table*}[!t]
        \renewcommand{\arraystretch}{1.3}
        \caption{Zero-Shot Classification on ESC-50 with Random Partition Strategy}
        \label{tab:ESC50_random}
        \centering
        \begin{tabular}{c||c|c||c|c||c|c}
            \hline
            \multirow{2}{*}{\bfseries Test Class Fold} & \multicolumn{2}{c||}{\bfseries WLE} & \multicolumn{2}{c||}{\bfseries GLE}
            & \multicolumn{2}{c}{\bfseries WLE+GLE}
            \\
            \cline{2-7}
            & mAP  & Top-1 & mAP            & Top-1          & mAP  & Top-1 \\
            \hline\hline
            Fold0 & 0.47 & 0.28  & 0.45           & 0.26           & 0.46 & 0.26  \\
            \hline
            Fold1 & 0.56 & 0.39  & \bfseries 0.61 & \bfseries 0.46 & 0.61 & 0.45  \\
            \hline
            Fold2 & 0.50 & 0.33  & 0.50           & 0.33           & 0.52 & 0.34  \\
            \hline
            Fold3 & 0.51 & 0.31  & 0.50           & 0.31           & 0.52 & 0.33  \\
            \hline
            Fold4 & 0.50 & 0.29  & 0.47           & 0.26           & 0.46 & 0.27  \\
            \hline
        \end{tabular}
    \end{table*}
    \begin{table*}[!t]
        \renewcommand{\arraystretch}{1.3}
        \caption{Averaged Performance on ESC-50 with Different Partition Strategies}
        \label{tab:ESC50_average}
        \centering
        \begin{tabular}{c||c|c||c|c}
            \hline
            \multirow{2}{*}{\bfseries Semantic Embedding} & \multicolumn{2}{c||}{\bfseries Category-based Strategy}
            & \multicolumn{2}{c}{\bfseries Random-based Strategy}
            \\
            \cline{2-5}
            & mAP            & Top-1          & mAP            & Top-1          \\
            \hline\hline
            WLE     & \bfseries 0.41 & \bfseries 0.23 & 0.51           & 0.32           \\
            \hline
            GLE     & 0.38           & 0.21           & 0.51           & 0.32           \\
            \hline
            WLE+GLE & 0.38           & 0.21           & \bfseries 0.52 & \bfseries 0.33 \\
            \hline
        \end{tabular}
    \end{table*}

    Overall, we obtain better results than random guess, which should give a mAP / \mbox{Top-1} of 0.29 / 0.10 in these experiments.
    The classification performance varies dramatically across different test class folds in both partition strategies regardless of label embeddings, which indicates that the partition strategy of training classes and test classes has a significant influence on zero-shot learning.
    Furthermore, we observe that all the three sets of label embeddings have best results on the ``Natural sounds" in the category-based strategy and the ``Fold1" fold in the random-based strategy.
    Among these label embeddings, WLE achieves the best mAP / \mbox{Top-1} of 0.52 / 0.33 on the ``Natural sounds" while GLE reaches the best mAP of 0.61 and \mbox{Top-1} of 0.46 on the ``Fold1" fold.

    On average, WLE achieves better performance in the category-based strategy with a mAP / \mbox{Top-1} of 0.41 / 0.23 while WLE+GLE in the random-based strategy reaches a mAP / \mbox{Top-1} of 0.52 / 0.33.
    Considering the performance on training data, we obtain an averaged training \mbox{Top-1} of 0.76 with WLE+GLE, 0.64 with WLE and 0.66 with GLE in the category-based strategy.
    In the random-based strategy, the averaged training \mbox{Top-1} is 0.81 with WLE+GLE, 0.76 with WLE and 0.83 with GLE\@.
    It indicates that over-fitting occurs with WLE+GLE in the category-based strategy and the partition strategy has an influence on the effectiveness of semantic embeddings.

    On the other hand, we notice that classification performance increases significantly in the random-based strategy in contrast to the category-based strategy.
    In the random-based strategy, sound classes from the same high-level category are distributed across training and test data.
    We conclude that classification performance can be improved by including classes that are semantically close to the test classes at training stage.
    Additionally, for the category-based strategy with WLE, we used a pre-trained VGGish to generate acoustic embeddings in our previous work\cite{refs:Xie2019ZeroShot} and obtained an averaged Top-1 of 0.26, which is better than the reported result (0.23) in Table~\ref{tab:ESC50_average}.
    It would be possible that the pre-trained VGGish already has acoustic information about the test classes, which leads to improved performance.

    \subsection{Zero-Shot Learning on AudioSet}
    \label{subsec:zero-shot-learning-on-audioset}

    The results of different semantic embeddings on two data settings are presented in Table~\ref{tab:AudioSet_result}.
    Overall, we obtain better results than random guess, which should give a mAP / Top-1 of 0.05 / 0.01 in these experiments.
    We discuss the effectiveness of different semantic embeddings in the following paragraphs.

    \begin{table*}[!t]
        \renewcommand{\arraystretch}{1.3}
        \caption{Zero-Shot Classification on the Extracted Subset of AudioSet via Different Semantic Embeddings with Two Data Settings}
        \label{tab:AudioSet_result}
        \centering
        \begin{tabular}{c|c||c|c||c|c}
            \hline
            \multirow{2}{*}{\bfseries Type} & \multirow{2}{*}{\bfseries Semantic Embedding} & \multicolumn{2}{c||}{\bfseries Setting 1}
            & \multicolumn{2}{c}{\bfseries Setting 2}
            \\
            \cline{3-6}
            &             & mAP            & Top-1          & mAP            & Top-1          \\
            \hline\hline
            \multirow{3}{*}{Label}    & WLE         & 0.15           & 0.06           & 0.15           & 0.06           \\
            \cline{2-6}
            & GLE         & 0.14           & 0.06           & 0.14           & 0.06           \\
            \cline{2-6}
            & WLE+GLE     & \bfseries 0.17 & \bfseries 0.08 & \bfseries 0.19 & \bfseries 0.08 \\
            \hline\hline
            \multirow{7}{*}{Sentence} & WSE         & 0.16           & 0.07           & 0.16           & 0.07           \\
            \cline{2-6}
            & GSE         & 0.15           & 0.06           & 0.15           & 0.06           \\
            \cline{2-6}
            & BSE         & 0.17           & \bfseries 0.08 & 0.14           & 0.07           \\
            \cline{2-6}
            & WSE+GSE     & 0.16           & \bfseries 0.08 & 0.17           & \bfseries 0.09 \\
            \cline{2-6}
            & WSE+BSE     & 0.14           & \bfseries 0.08 & 0.17           & 0.08           \\
            \cline{2-6}
            & GSE+BSE     & \bfseries 0.18 & \bfseries 0.08 & 0.17           & \bfseries 0.09 \\
            \cline{2-6}
            & WSE+GSE+BSE & 0.16           & \bfseries 0.08 & \bfseries 0.18 & \bfseries 0.09 \\
            \hline\hline
            \multirow{5}{*}{Hybrid}   & WLE+WSE     & 0.18           & 0.07           & 0.18           & 0.07           \\
            \cline{2-6}
            & GLE+GSE     & 0.17           & 0.07           & \bfseries 0.21 & \bfseries 0.11 \\
            \cline{2-6}
            & WLE+BSE     & \bfseries 0.21 & 0.10           & 0.19           & 0.10           \\
            \cline{2-6}
            & GLE+BSE     & 0.20           & 0.11           & 0.20           & \bfseries 0.11 \\
            \cline{2-6}
            & WLE+GLE+BSE & 0.20           & \bfseries 0.12 & 0.20           & \bfseries 0.11 \\
            \hline
        \end{tabular}
    \end{table*}

    \textbf{Comparison of label embeddings}.
    We first compare different label embeddings.
    For WLE and GLE, similar mAPs (around 0.15) and accuracies (about 0.06) are obtained in both data settings.
    By concatenating WLE and GLE, better results are reached with a mAP / \mbox{Top-1} of 0.17 / 0.08 in \mbox{Setting 1} and 0.19 / 0.08 in \mbox{Setting 2}, respectively.
    It shows that classification performance is improved by concatenating individual label embeddings.

    \textbf{Comparison of sentence embeddings}.
    We experiment with seven sets of sentence embeddings: individual embeddings (i.e., WSE, GSE and BSE), which are generated with individual language models, and their concatenations (i.e., WSE+GSE, WSE+BSE, GSE+BSE, and WSE+GSE+BSE).
    For the sake of comparing individual embeddings and their concatenations, we average the results across these embeddings.
    For individual embeddings, an averaged mAP / Top-1 of 0.16 / 0.07 is achieved in \mbox{Setting 1}, and 0.15 / 0.07 in \mbox{Setting 2}.
    In contrast, their concatenations reach an averaged mAP / Top-1 of 0.16 / 0.08 in \mbox{Setting 1}, and 0.17 / 0.09 in \mbox{Setting 2}.
    It shows that classification accuracy is improved by concatenating individual sentence embeddings.

    \textbf{Comparison of hybrid embeddings}.
    For comparing hybrid embeddings, we divide them into two groups according to their concatenated sentence embeddings (i.e., WSE/GSE and BSE).
    For hybrid embeddings with WSE/GSE (i.e, WLE+WSE and GLE+GSE), an averaged mAP / Top-1 of 0.18 / 0.07 is achieved in \mbox{Setting 1}, and 0.20 / 0.09 in \mbox{Setting 2}.
    On the other hand, we observe that hybrid embeddings with BSE (i.e., WLE+BSE, GLE+BSE and WLE+GLE+BSE) achieve an averaged mAP / Top-1 of 0.20 / 0.11 in both data settings, which is better than those of hybrid embeddings with WSE/GSE\@.

    Similarly to label/sentence embeddings, it shows that concatenating semantic embeddings generated with different language models helps improve classification performance.
    Since different language models can learn different aspects (in Section~\ref{subsec:semantic-embeddings}) of semantic information from textual labels and sentence descriptions, concatenated embeddings will contain richer semantic information than individual embeddings, which results in an improvement on the performance.

    \textbf{Comparison of label/sentence/hybrid embeddings}.
    First, we compare individual label embeddings (i.e., WLE and GLE) and their hybrid embeddings (i.e., WLE+WSE/WLE+BSE and GLE+GSE/GLE+BSE).
    The results in Table~\ref{tab:AudioSet_result} show that classification performance is improved for WLE and GLE by concatenating them with individual sentence embeddings.
    Secondly, we compare individual sentence embeddings (i.e., WSE, GSE and BSE) and their hybrid embeddings (i.e., WLE+WSE, GLE+GSE, and WLE+BSE/GLE+BSE).
    We observe that WLE+WSE reaches a better mAP (0.18) than WSE (0.16) but has the same Top-1 (0.07) in both settings.
    For GSE, classification performance is improved by combining it with GLE (i.e., GLE+GSE) in both settings.
    Particularly, GLE+GSE achieves a mAP / Top-1 of 0.21 / 0.11 in \mbox{Setting 2}.
    Similarly, we obtain better results with the hybrid embeddings containing BSE (i.e., WLE+BSE and GLE+BSE) than the individual BSE\@.
    Therefore, we conclude that classification performance can be improved by concatenating individual label embeddings and sentence embeddings.

    \begin{table}[!t]
        \renewcommand{\arraystretch}{1.3}
        \caption{Averaged Performance across Semantic Embeddings on Two Data Settings of the Extracted Subset of AudioSet}
        \label{tab:AudioSet_average}
        \centering
        \begin{tabular}{c||c|c||c|c}
            \hline
            \multirow{2}{*}{\bfseries Type} & \multicolumn{2}{c||}{\bfseries Setting 1} & \multicolumn{2}{c}{\bfseries Setting 2} \\
            \cline{2-5}
            & mAP            & Top-1          & mAP            & Top-1          \\
            \hline\hline
            Label    & 0.15           & 0.07           & 0.16           & 0.07           \\
            \hline
            Sentence & 0.16           & 0.08           & 0.16           & 0.08           \\
            \hline
            Hybrid   & \bfseries 0.19 & \bfseries 0.09 & \bfseries 0.20 & \bfseries 0.10 \\
            \hline
        \end{tabular}
    \end{table}

    Furthermore, we study the overall effectiveness of these three types of semantic embeddings by calculating their averaged mAP / Top-1 in Table~\ref{tab:AudioSet_average}.
    For label embeddings, we obtain an averaged mAP / Top-1 of 0.15 / 0.07 in \mbox{Setting 1}, and 0.16 / 0.07 in \mbox{Setting 2} across WLE, GLE and WLE+GLE\@.
    For the seven sets of sentence embeddings, an averaged mAP / Top-1 of 0.16 / 0.08 is achieved in both data settings.
    Meanwhile, the hybrid embeddings reach an averaged mAP / Top-1 of 0.19 / 0.09 in \mbox{Setting 1}, and 0.20 / 0.10 in \mbox{Setting 2}.
    It shows that sentence embeddings have better averaged accuracies than label embeddings while hybrid embeddings have the best performance.
    A possible interpretation for this would be that sentence descriptions convey richer semantic information than textual labels while their combination includes all their semantic information, which results in an improvement on classification performance.

    \textbf{Comparison of data settings}.
    Table~\ref{tab:AudioSet_average} shows that similar averaged results (i.e., mAP and Top-1) are observed in both data settings for different semantic embeddings.
    It would be probably due to the sound classes in these AudioSet class folds being comprehensive enough.

    \textbf{McNemar's test}.
    We apply McNemar's test\cite{refs:Kavzoglu2013Anassessment} to analyze statistical significance of the differences between the results of semantic embeddings.
    For the sake of simplicity, we compare WLE+GLE+BSE and GLE+BSE, which achieve the best Top-1 and the second best Top-1 in \mbox{Setting 1}.
    With the contingency Table~\ref{tab:contingency_table}, we obtain a McNemar's test statistic of 52.05 and a p-value of \mbox{5.41e$-$13}, which shows a significant difference between the results of WLE+GLE+BSE and GLE+BSE\@.

    \begin{table}[!t]
        \renewcommand{\arraystretch}{1.3}
        \caption{Contingency Table for Semantic Embeddings WLE+GLE+BSE and GLE+BSE in AudioSet}
        \label{tab:contingency_table}
        \centering
        \begin{tabular}{c|c|c|c}
            \hline
            \multicolumn{2}{c|}{\bfseries Correct / Incorrect} & \multicolumn{2}{c}{\bfseries WLE+GLE+BSE} \\
            \cline{3-4}
            \multicolumn{2}{c|}{\bfseries Predictions} & \blap{correct} & \blap{incorrect} \\
            \hline
            \multirow{2}{*}{\bfseries GLE+BSE} & \blap{correct}   & 1854 & 381   \\
            \cline{2-4}
            & \blap{incorrect} & 609  & 18533 \\
            \hline
        \end{tabular}
    \end{table}

    \textbf{Summary}.
    The experimental results show that label/sentence embeddings of sound classes, such as WLE/GLE and WSE/GSE/BSE, are useful for zero-shot learning in audio classification.
    Classification performance can be improved by concatenating individual label/sentence embeddings generated with different language models, such as WLE+GLE and WSE+GSE\@.
    With hybrid embeddings (e.g., WLE+BSE, GLE+BSE), which are obtained by concatenating individual label embeddings and sentence embeddings, the results are improved further.

    \subsection{Discussion}
    \label{subsec:discussion}

    So far, zero-shot learning for audio classification has received relatively little attention, for example, in comparison to computer vision research.
    When formulated as a multi-label classification problem with some feature inputs, these two fields are very similar.
    On the other hand, there are some factors that make the fields different.
    In computer vision, zero-shot learning has been studied with even larger datasets (e.g., ImageNet with over 21,000 classes\cite{refs:Deng2009ImageNet}), in comparison to existing works (e.g., 1,126 classes used in\cite{refs:Choi2019ZeroShot} and 521 classes in this work) in the audio field.
    With the fine-grained classes in ImageNet, Xian et al.\cite{refs:Xian2018ZeroShot} conducted an in-depth study of zero-shot learning with respect to the semantic similarity, hierarchy and popularity between image classes.
    Meanwhile, a wider set of semantic information has been explored for zero-shot learning in computer vision.
    For instance, human-defined visual attributes have been widely used as the semantic information of visual classes\mbox{\cite{refs:Lampert2014Attribute, refs:RomeraParedes2015An, refs:Akata2016Label, refs:Xian2018ZeroShot}}.
    Class hierarchies have also been studied for zero-shot learning in\mbox{\cite{refs:Fergus2010Semantic, refs:Akata2016Label}}.
    Another factor is about the intrinsic properties of classes in these two fields, which are also present in supervised classification.
    For example, in audio, multiple sound sources are often present simultaneously, forming a complex mixture of sounds.
    Visual objects may be partly occluded, but not usually distorted by other sources.
    Furthermore, the presence of visual objects in static image can be considered to be binary (present or not), whereas the activities of sound classes are almost always time-varying.

    \section{Conclusion}
    \label{sec:conclusion}

    In this paper, we present a zero-shot audio classification method via semantic embeddings that are learned from either textual labels or sentence descriptions of sound classes.
    The experimental results on \mbox{ESC-50} show that classification performance is significantly improved by involving sound classes that are semantically close to the test classes in training.
    Furthermore, we explore zero-shot learning on a large-scale audio dataset, which contains 521 sound classes extracted from AudioSet.
    We demonstrate that both label and sentence embeddings are useful for zero-shot audio classification.
    Compared with individual label/sentence embeddings, concatenating embeddings generated with language models improves the results.

    There is plenty of room to explore for future work.
    For instance, the pre-trained language models can be fine-tuned with sound-specific text corpora for generating high-quality semantic embeddings of sound classes.
    To improve label and sentence embeddings, word embeddings can be aggregated using advanced methodologies from natural language processing literature instead of averaging them.
    We also notice that different semantic embeddings result in learning variable volumes of parameters in the bilinear compatibility function, which would have an effect on classification performance.
    Future studies on semantic embeddings should take these into account.
    On the other hand, instead of averaging VGGish-generated embeddings of non-overlapping segments, acoustic embeddings can be improved by generating them from the entire audio clips with advanced models (e.g., attention models).
    Meanwhile, training acoustic/language embedding models jointly with the bilinear compatibility learning framework would be beneficial to zero-shot learning.
    Furthermore, rather than a simple linear acoustic-semantic projection, nonlinear ones should be investigated in the future.

    \ifCLASSOPTIONcaptionsoff
    \newpage
    \fi

    \bibliographystyle{IEEEtran}
    \bibliography{IEEEabrv,refs}

\end{document}